\documentclass[preprint,aps,endfloats]{revtex4}

\usepackage{graphicx}
\usepackage{multirow}
\usepackage{dcolumn}
\usepackage{bm}
\usepackage{amssymb}

\begin{document}

\preprint{APS/xxxxx}

\title{Silicon dry oxidation kinetics at low temperature in the nanometric range: modeling and experiment}

\author{Christophe Krzeminski, Guilhem Larrieu, Julien Penaud, Evelyne Lampin and  Emmanuel Dubois}
\affiliation{Institut d'Electronique, de Micro\'electronique et de Nanotechnologies, UMR CNRS 8520-D\'epartement ISEN, Avenue Poincar\'e, Cit\'e Scientifique, BP 60069, 59652 Villeneuve d'Ascq Cedex, France}

\email{christophe.krzeminski@isen.fr}

\vspace{10 cm}

\begin{abstract}
Kinetics of  silicon dry oxidation are investigated theoretically and experimentally at low temperature in the nanometer range where the limits of the Deal and Grove model becomes critical. Based on a fine control of the oxidation process conditions, experiments allow the investigation of the  growth kinetics of  nanometric oxide layer. The theoretical model is formulated using a  reaction rate approach. In this framework, the oxide thickness is estimated with the evolution of the various species  during the reaction.   Standard oxidation models and the reaction rate approach  are confronted with these experiments. The interest of the reaction rate approach to improve  silicon oxidation modeling in the nanometer range is clearly demonstrated.
\end{abstract}


\maketitle

%

\section{Introduction}
An important challenge imposed by CMOS downscaling is the growth of ultra-thin oxides of silicon of high quality, with a tight thickness control and a good uniformity \cite{ITRS}. Very important progress have been done during the last decade to improve the structural properties and electrical limits of ultra-thin ($<4$ nm) silicon oxide and oxynitride \cite{Green01}. On the other side,  the  modeling  of ultra-thin silicon oxide growth remains a difficult issue  for the microelectronics industry.

To date, the seminal work of Deal and Grove  remains the main approach used in process simulators to describe the oxide growth \cite{deal65}. Assuming  a steady state  reaction between  molecular oxygen and silicon, Deal and Grove deduced that the oxide growth can simply be described by the following equation :

\begin{equation}
X= \frac{A}{2} \cdot \left [ \sqrt{1+\frac{t+\tau}{A^{2}/4B}}-1 \right ]
\end{equation}

where the oxide thickness $X$  is described by a linear-parabolic relationship as a function of the oxidation time $t$. The term $\frac{B}{A}$ characterizes the initial linear rate growth. The  parabolic rate constant $B$ governs the diffusion limited regime. An initial time offset $\tau$ is necessary to take into account the presence of an initial thick oxide layer. The success of this model can be explained by its ability to describe both dry and wet oxide in the case of thick oxide. Another major achievement is the introduction  by Kao et al. \cite{Kao87a} of the influence of strain effects in the Deal and Grove approach. Using this generalized Deal and Grove approach, Kao et al.  explained the retardation effect observed for the oxidation of  curved surfaces \cite{Kao87b}.

However, this model suffers from strong limitations such as  the failure to describe the fast initial growth regime  for dry oxidation \cite{Massoud85a} or the difficulties to extend it to oxynitridation process \cite{Dimitrijev96}. Much research work has been dedicated to elucidate the breakdown of the Deal \& Grove model in the ultra-thin regime. The idea  was to correct the Deal and Grove model by adding some empirical terms to the growth rate \cite{Massoud85b, Han87}. However, the physical origin of these terms is subject to discussion. 
 
An alternative approach has been proposed by Wolters et al. who considered that mobile ionic species are responsible of  the silicon  oxidation process \cite{Wolters89a, Wolters89b}. Ionic transport and induced electric field considered in this model could explain some of the experimental phenomena observed in silicon oxidation such as the rate dependence on the orientation or the cross-over effects \cite{Wolters93}. But  the role of charged species in the oxidation mechanism is still debated even, for  more fundamental simulations such as  density functional theory \cite{Stoneham01, Bongiorno04}. Recently, a new alternative linear parabolic rate equation  has also been proposed by Watanabe et al. \cite{Watanabe06} without the rate-limiting step of the oxidation reaction but assuming that diffusivity is suppressed in a strained oxide near the silicon  interface.

From the above discussion, a revision of the  Deal and Grove model is clearly necessary to improve process simulation. An  ideal model for the microelectronics industry must  give access to : (i) a fundamental and quantitative understanding of the interaction between oxygen, silicon and the standard dopant at the Si/SiO$_{2}$  to better describe  segregation effects \cite{kim05, kohli05} or to estimate the dose-loss of dopant, (ii) the amount of  silicon interstitials at the interface (iii) the amount of nitrogen incorporated during a more complex  oxynitridation process (NO, N$_{2}$O gas) to estimate the gate leakage current \cite{Muraoka03}. The development of such  model still represents a big challenge. Moreover, a  multiscale modeling of silicon oxidation  \cite{Bongiorno04} will probably be mandatory  to simulate the growth of ultimate oxides ( $\sim$ 1 nm) fabricated by advanced oxidation  process   such as plasma nitridation step \cite{Pan05}.

In this paper, the  modeling of dry oxidation kinetics at  low temperature in the nanometer range is investigated.  This paper is organized as follows. In section II,  the reaction rate formalism is presented. The calibration method and some preliminary results are also discussed. Experiment at low temperature are  described in section III.  The two standard models (Deal and Grove \cite{deal65}, Massoud \cite{Massoud85a,Massoud85b}) and the reaction rate approach are confronted with the experimental data in section IV. Most important conclusions are summarized in section V.

\section{The reaction rate approach}

\subsection{Introduction}

An interesting study in  silicon oxidation modeling has been published by Almeida et al.   \cite{Almeida00, Baumvol99}. The main idea  of this work is to analyze the  main assumptions of the Deal and Grove model and to propose a  more rigorous approach. The first approximation made by Deal and Grove  is to consider that the reaction  strictly takes place at the interface. The second approximation is the steady state regime which imposes the a balance  between oxidizing species entering  the SiO$_{2}$ surface, the number of molecules diffusing into the oxide  and the  molecules reacting at the Si/SiO$_{2}$ interface.  Thus the concentration of the oxidant species is not time dependent. As  outlined in an very elegant way by Almeida et al., these  assumptions are certainly the key  to explain the  limitations  of the Deal and Grove model in the ultra-thin regime for dry oxidation. Moreover  these assumptions limits the development of more sophisticated treatment of the oxidation mechanism.

\subsection{The model} 

\begin{figure}
\includegraphics[width=6cm]{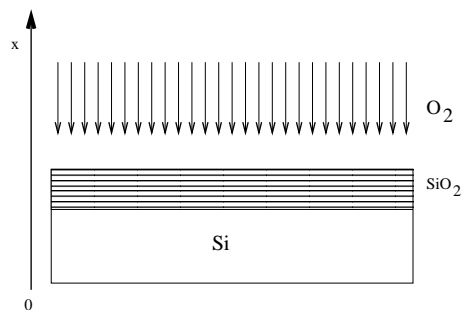}
\caption{Experimental system under investigation. Oxygen diffuses and reacts with silicon and creates silicon dioxide. A very  fine 1D mesh  is defined to simulate the evolution of the various concentration.
\label{fig:figscheme}}
\end{figure}

In this paper,  a  reaction rate approach similar to that of Almeida et al.  \cite{Almeida00, Baumvol99} has been adopted for the modeling of our experimental kinetic at low temperature.  The main diffusing species is  assumed to be molecular oxygen (O$_{2}$) as demonstrated by  several isotopic experiments \cite{Rochet84}. The two other species are pure silicon (Si) and silicon dioxide (SiO$_{2}$). The reaction of oxidation is simply given by:

\medskip

\begin{center}
\begin{tabular}{ccc} 
&$K$&\\
Si+ O$_{2}$ & $\Longrightarrow$ & SiO$_{2}$
\label{react}
\end{tabular}
\end{center}

where $K$ is the reaction rate. Keeping in mind that the investigation focuses on  the nanometric regime,  a $\left[100\right]$ silicon surface associated with a one dimensional system of coordinates is considered  (see Figure \ref{fig:figscheme}). The area to be oxidized is viewed as an assembly of silicon monolayers. The grid is defined to match as fine as possible this assembly of planes. This means that the  vertical mesh  step  corresponds to the distance between two silicon planes (1.35 \r{A}). The oxygen flux in the furnace is perpendicular to the silicon interface. Since the model involves an  oxidation mechanism in the nanometer range, the growth of the film is restricted to the vertical direction.

The relative concentration $n_{j}$ of the  species $j$ (where $j$= Si, O$_{2}$, and SiO$_{2}$) is defined as follow :

\medskip
\begin{equation}
n_{j}(x,t)=\frac{C_{j}(x,t)}{C_{j}^{0}}
\label{one}
\end{equation}
\medskip

where $C_{j(x,t)}$ corresponds to the concentration in the plane  in units of number of atoms per unit surface, and $C_{j}^{0}$ is the maximum possible  concentration in the plane. $n_{j}(x,t)$ can be viewed  as a  layer coverage. For example, a value of 1 for $n_{Si}$, means that the coverage of the layer is complete and corresponds to a concentration of  0.91$\times 10^{15}$ at/cm$^{2}$ of silicon.

In order to setup the system of equations, the evolution of each species is described. During the oxidation process, the  molecular oxygen  diffuses in the silicon dioxide, reaches the silicon and reacts with it. Considering  the diffusivity of the molecular oxygen  $D$ and the reaction rate constant $K$, the evolution of molecular oxygen can be given by:

\begin{equation}
\frac{\partial n_{O_{2}}}{\partial t}=\nabla (D\cdot \nabla n_{O_{2}})-K\cdot n_{O_{2}}\cdot n_{Si}
\label{eq:oxi1}
\end{equation}

In this expression, the reaction  region is defined by the overlap between the  oxygen and silicon species. Next, silicon consumption has is described by:
\begin{equation}
\frac{\partial n_{Si}}{\partial t}=-K \cdot n_{O_{2}}\cdot n_{Si}
\label{eq:oxi2}
\end{equation}

To simplify the previous equation, the fact that some  silicon interstitials  generated during oxidation  \cite{Skarlatos99} has been neglected. The conservation law of the different species allows  the description of the silicon dioxide creation:

\begin{equation}
\frac{\partial n_{SiO_{2}}}{\partial t}=K\cdot n_{O_{2}}\cdot n_{Si}
\label{eq:oxi3}
\end{equation}

The mathematical formulation of the three steps (diffusion of  molecular oxygen, reaction with silicon, and creation of silicon dioxide) leads finally  to a  system of three coupled  equations (Eqs \ref{eq:oxi1}, \ref{eq:oxi2}, \ref{eq:oxi3}). This system is  numerically solved using a Cranck-Nicolson scheme \cite{theart}.

\subsection{Boundary conditions}

To complete the mathematical description of the oxidation mechanism, boundary conditions need to be refined. For  oxygen, it is necessary to  evaluate  the concentration of oxygen molecules that lies on the silicon surface. In this study,  this concentration is estimated by simple physical considerations.

Considering that the oxygen   is described by the ideal gas law, the number of  oxygen molecules in the furnace chamber can be expressed as a function of the gas pressure $P$ and temperature $T$. Assuming that an ideal monolayer of oxygen molecules lies on top of the wafer surface, the density is :

\medskip
\begin{equation}
C_{O_{2}}=\frac{N_{O_{2}}}{S}=\frac{N_{A} P h}{RT}
\label{fourteen}
\end{equation}
\medskip

where $N_{A}$ is the  Avogadro number, $R$ the universal gas constant,  $S$ is the surface of the lattice and $h$ the height of the  oxygen layer. If  the maximum density $C^{max}_{O_{2}}$ occurs when all the molecules form a square lattice of $a=3$ \r{A} spacing:

\medskip
\begin{equation}
C^{max}_{O_{2}}=\frac{1}{S_{min}}\simeq 1.0 \times 10^{15} at/cm^{2}
\label{fithteen}
\end{equation}
\medskip

The oxygen layer coverage on the surface can be expressed by the following expression  :

\medskip
\begin{equation}
n^{0}_{O_{2}}=\frac{C_{O_{2}}}{C^{max}_{O_{2}}}=\frac{N_{A}a^{2}h P}{RT}=0.161\cdot \frac{P}{T}
\label{sixteen}
\end{equation}
\medskip

which further simplifies to the  last expression of Eq. \ref{sixteen}, if the  height of the layer is about $h=3$ \r{A}.

\begin{figure}
\includegraphics[width=7cm]{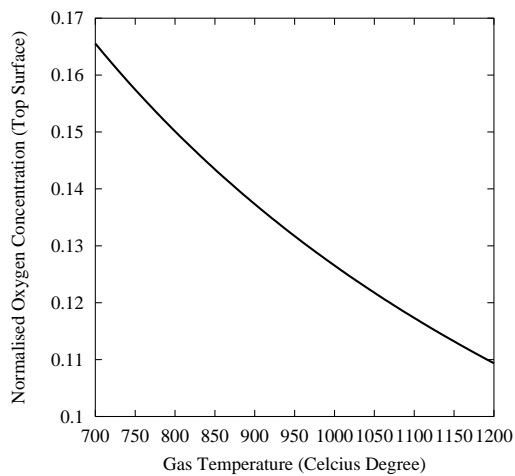}
\caption{Variation of the layer coverage of oxygen with temperature. \hspace{5cm}
\label{fig:figcoveragetemp}}
\end{figure}

The layer coverage of oxygen at the interface depends on both   the partial pressure  and the temperature.  Figure \ref{fig:figcoveragetemp} represents the variation of this coverage with the temperature. 

\

\subsection{Calibration}

A rough estimation of the the oxygen  concentration at the interface can be deduced using  simple physical arguments. In order to  estimate the  value for  the reaction rate $K$ and for the diffusivity  $D$, a  calibration step  has been undertaken on  experimental  kinetics. The large experimental oxidation database of Massoud (measured with an in-situ ellipsometer) has been chosen \cite{Massoud85a}. A numerical  optimization method based on simulated annealing  has been undertaken   \cite{Goffe94}. This method explores   the  whole solution space for the parameters (here $K$, $D$) to reach a global minimum for the error function. The probability of jump between  different parameter values is proportional to Boltzmann probability distribution. At the beginning of the  calibration step,  the temperature $T_{B}$ (for the Boltzmann distribution) is  high to explore all the solution space and then  $T_{B}$ is  cooled down to limit the variation of the parameters in order to  reach a minimum. The calibration procedure   used is :

\begin{itemize}
\item{Start with an initial value of $K$ and $D$ and with an important value for the control parameter $T_{B}$ }
\medskip

\item{Calculate the theoretical oxide thickness $X_{th}$ at the different experimental points}
\medskip

\item{Estimate the error function $E$ defined by the difference between the theoretical estimations and the different experimental thickness of the kinetics:}

\begin{equation}
E=\sqrt{\sum_{i=1}^{Nexp} \alpha_{i} \times \big [ X_{th}(t_{i})-X_{exp}(t_{i}) \big ]^{2}}
\end{equation}
\medskip

Where $Nexp$  is the number of the experimental points of the oxidation kinetics. $\alpha_{i}$ are weighting coefficients defined such as to enforce the algorithm to minimize principally the error function  for the lowest oxide thickness.

\item{Modify the value of $K$ and $D$ to minimize the error function $E$. According to the Boltzmann  probability associated with E, cool $T_{B}$}.

\end{itemize} 

At the end of an important  number of iteration steps (100000), the value of the two parameters minimizing  the error function  is obtained.

\subsection{Calibration results}

\begin{figure}
\includegraphics[width=7cm]{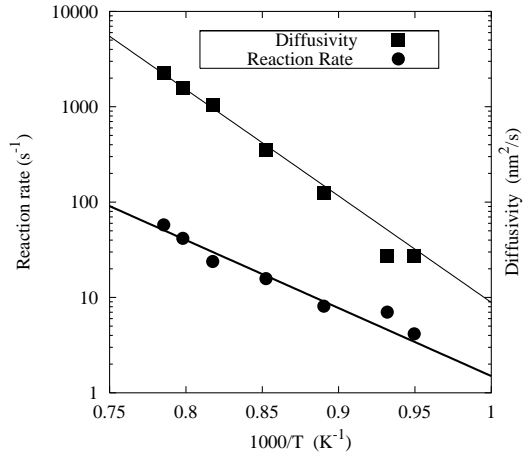}
\caption{Reaction rate (K) and  diffusivity coefficient (D)  obtained by the calibration step for  the different experimental  kinetics of  Massoud  et al.  for different  temperatures. The Arrhenius  laws  are also reported.
\label{fig:fig_k_d}}
\end{figure}

The parameters given by the optimization step to adjust the  Massoud experimental data are reported in Table \ref{table:oxid_one}. In order to extract a  physical law for the variation of the two parameters,  the different values have been calibrated again  to match an Arrhenius law. As shown in  figure \ref{fig:fig_k_d}, both the diffusivity and the reaction rate parameter agree well with an  Arrhenius plot, which leads to the following law:

\begin{equation}
D=D_{0} \cdot \exp{ \bigg [-\frac{E_{D}}{k_{B}T} \bigg  ]}
\end{equation}

The  energy of activation is found to be $E_{D}=2.22$ eV and the prefactor $D_{0}=1.291 \times 10^{11}$ nm$^{2}$/s. For the reaction-rate, we obtain similarly an Arrhenius law  :

\begin{equation}
K=K_{0} \cdot \exp{  \bigg  [-\frac{E_{K}}{k_{B}T} \bigg ]}
\end{equation}

with $K_{0}=2.022 \times 10^{7} s^{-1}$ and $E_{K}=1.42 $eV. It is worth noting that the activation energy and reaction rate are physically reasonable and  close to  published numbers  \cite{Almeida00,Han87,kim96}. The calibrated activation energy for the diffusivity is in agreement with the experimental diffusion energy in silicon (2.42 eV) \cite{kim96} and  with {\it ab-initio} estimations (2.3-2.5 eV). The energy of activation (1.42 eV) for the reaction rate is not so different  from the energy to break an Si-Si bonds (1.86 eV) \cite{Pauling60}.

\subsection{Preliminary evaluation}
\begin{figure}
\includegraphics[width=7cm,angle=0]{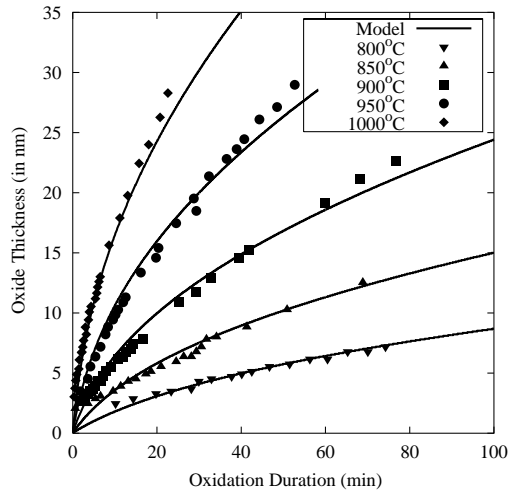}
\caption{Experimental oxidation kinetics  of Massoud (Points) et al. \cite{Massoud85a}  and calculated with our model (lines).
\label{fig:figmassoud}}
\end{figure}

\begin{figure}
\includegraphics[width=7cm,angle=0]{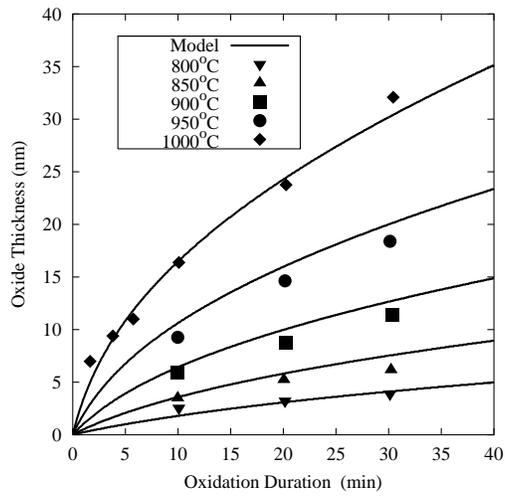}
\caption{Experimental kinetics of  Chao et al. \cite{Chao91} for dry oxidation (points) and the kinetics predicted by the new approach (lines).
\label{fig:fig_chao}}
\end{figure}

A primary consolidation of the present model has been performed through a comparison with published oxide thickness measured by ellipsometry. Figure \ref{fig:figmassoud} compares the prediction of our modeling approach with  the kinetics of  Massoud et al. at atmospheric pressure  from 800$^{\circ}$C to 1000$^{\circ}$C. Each sample has an 1.0 nm initial oxide prior to oxidation.  Thanks to the calibration step,  an excellent agreement between the experimental data and the model is obtained. The oxidation rate for the different temperature is clearly well described both in the linear and the parabolic regime. No clear loss of predictivity is observed when using the analytical law for the reaction rate and the diffusivity rather than each pair of calibrated data. Next,  the model is compared with the experiments made by Chao et al. \cite{Chao91} at atmospheric pressure in the nanometer range. The corresponding experimental data shown in the figure \ref{fig:fig_chao} were obtained with a multi-angle incident ellipsometer. Samples have a native oxide of 1.6-2nm covering the substrate. Similar conclusions can be drawn  again considering that the variation of the oxidation growth rate  is well described by the model for experimental conditions close to those used by Massoud et al.

\begin{figure}
\includegraphics[width=7cm,angle=0]{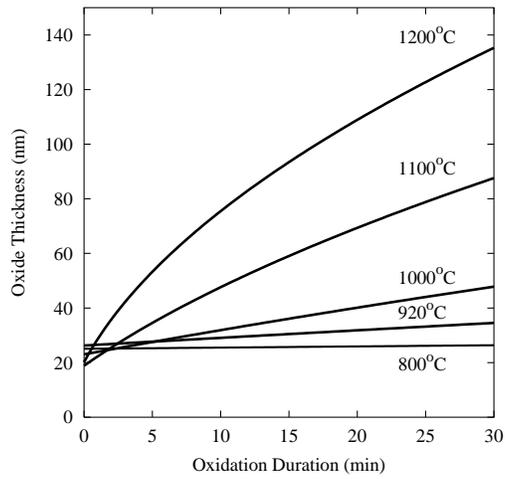}
\caption{Original dry oxidation kinetics predicted by  Deal and Grove \cite{deal65} obtained  with  the original parameters. 
\label{fig:fig_deal1}}
\end{figure}

\begin{figure}
\includegraphics[width=7cm,angle=0]{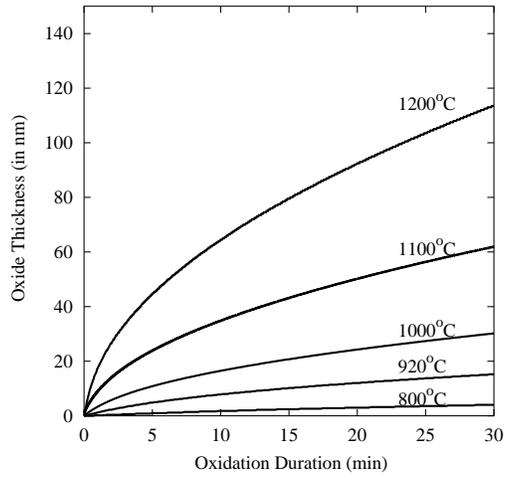}
\caption{Oxidation kinetics predicted by the reaction rate approach for the same conditions (temperature, pressure).
\label{fig:fig_react_deal}}
\end{figure}

Finally, a comparison between our model and the original kinetics obtained by the Deal and Grove model has been performed. The main objective is to check the validity of the reaction rate approach to describe correctly the oxide growth in the thick regime. The comparison between the two models is of interest. Figure \ref{fig:fig_deal1} presents the classical Deal and Grove kinetics simulated with the {\it original} parameters published in 1965 \cite{deal65}. The breakdown of Deal and Grove model to describe  dry oxide growth below 20 nm is clearly emphasized. Almost no oxide growth is   observed for a temperature near  800$^{\circ}$C. The kinetics obtained by the reaction rate approach with the same temperature and pressure  is represented in the figure \ref{fig:fig_react_deal}. The differences between the two models is obvious in the ultra-thin regime. However the reaction rate approach is also able to describe the oxide growth in the thick regime. The growth rates predicted by the two models for  various temperatures are in relatively good agreement. Even if the initial thickness considered in the two models is strongly different, the amount of oxide grown is almost identical a  difference that does not exceed 5 $\%$ for 30 minutes at 1200$^{\circ}$C.

\section{Dry oxidation at low temperature}

A complementary experimental work has been undertaken to validate the model.  In order to reach the nanometer range with a conventional furnace, the oxidation is often performed by reducing the pressure of the growth ambient  \cite{bidaud01, ludsteck04}, by diluting the oxidant gas with nitrogen \cite{Bhat01a}, or by lowering the temperature \cite{Irene76}. Oxidation with low thermal budget is an interesting solution. This point is illustrated by the study of Bhat et al. who propose the growth of ultra-thin oxides of silicon at low temperature [600$^{\circ}$C-700$^{\circ}$C] by wet oxidation \cite{Bhat01b}. In the present case, dry oxidation in a classical furnace is performed at low  temperatures (725$^{\circ}$C, 750$^{\circ}$C) to fabricate oxide in the [1.5-4 nm] range.

\subsection{Oxidation experiments}

\begin{figure}
\includegraphics[width=10cm]{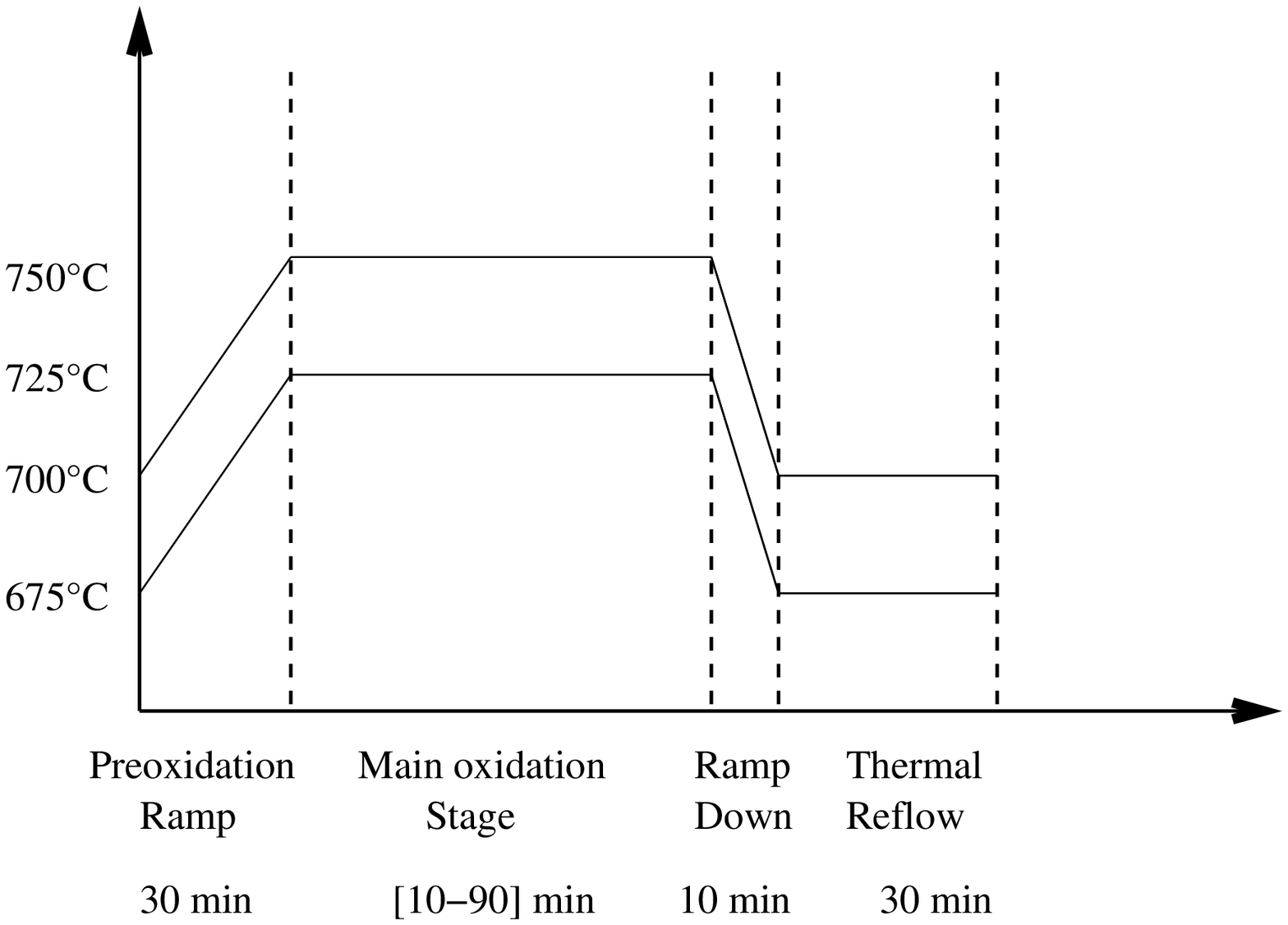}
\caption{The two thermal oxidation cycles at (725$^{\circ}$C, 750$^{\circ}$C) used in our oxidation experiments. Each thermal cycle is composed of a pre-oxidation ramp, an isothermal oxidation step and a thermal reflow.
\label{fig:figfurnace}}
\end{figure}

Low oxidation temperature is not sufficient to achieve oxide layers with interesting structural properties in the nanometer range. Special care has to be given on the design of the thermal process to achieve a good homogeneity. Our experiments were performed on [100] silicon substrates (p type doped 5$\times$ 10$^{15}$ at/cm$^{3}$) that previously received a RCA cleaning and 1$\%$ HF dip. The oxide growth was carried out in a conventional furnace at atmospheric pressure. The oxidation process is divided into four stages as shown in figure \ref{fig:figfurnace}.

 The sample introduction is performed at a temperature lower by 50$^{\circ}$C than the main  temperature step. Next, a temperature ramp is applied to reach the desired temperature  during 30 minutes. The ambient is composed by  a main  flux of nitrogen (2 L/min)  and a small amount of oxygen (0.2 L/min). During this ramp, a very thin layer of silicon dioxide is created on top of the wafer before the start  of the isothermal part of the oxidation step. An oxide of 1.53 nm (resp. 1.96 nm) is grown for a preoxidation ramp of  30 minutes at 725$^{\circ}$C (resp. 750 $^{\circ}$C)  is created during this step. The thickness grown during the first  part of the thermal cycle is nearly half of the maximum thickness range. However, this step is critical regarding the oxide thickness homogeneity. The preoxidation ramp has a clear impact on homogeneity  which is  less than 2 $\%$ on the whole wafer for the oxidation process at 725$^{\circ}$C. Performing the same  process with an inert ambient during the ramp-up leads to a strong increase of the oxide thickness dispersion up to 16  $\%$.

Finally, the oxidation stage is realized under an enhanced oxygen flux of (2 L/min). As shown in table II, the oxide growth is slow and the expected final thickness can be precisely controlled by the duration of the main oxidation stage since the reaction rate is very limited. The final step consists in an inert nitrogen atmosphere for a ramp down during 10 min. and by a thermal reflow of 30 min. No  growth is expected during this phase. The  main objective of the final  process step is to improve the electrical properties of the oxide film and to reduces the amount of interface defects.

\subsection{Ellipsometry and TEM measurements}

\begin{figure}
\includegraphics[width=7cm]{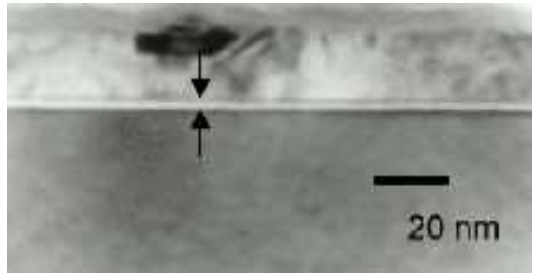}
\caption{XTEM analysis of an oxide layer of 19 \r{A} grown at 725$^{\circ}$C during 10 minutes. \hspace{3cm}
\label{fig:figtem}}
\end{figure}

Oxide thickness has  been measured  by a spectroscopic ellipsometer. In order to validate the ellipsometry  measurements, some TEM analysis were performed. Two oxide layers grown  with a preoxidation ramp  at 725$^{\circ}$C during 10 min and 30 min were analyzed. A  polysilicon capping  layer simulate the gate stack. The TEM analyzes (figure \ref{fig:figtem}) shows that a thin and uniform oxide layer. The following table (table \ref{table:oxid_two}) compare the thickness measurements given by the two methods. The good agreement between the ellipsometer and TEM measurements validate the ellipsometric method even if it has already  been observed that for thickness below 2.5 nm, deviations become more important  \cite{Marel04}.

\section{Simulation of oxide growth at low temperature.}

\subsection{Introduction}

The main objective of this section is to compare our experimental kinetics with the two standard models (Deal and Grove and Massoud) and the reaction rate approach. It has been previously verified that the reaction rate approach  is able describe the kinetics in the case of thick oxides. However some questions still require clarification: i) is finally the reaction rate approach more adapted to describe the silicon oxidation in the nanometer regime ? ii) Is it possible to describe the oxidation in the low temperature regime where the oxidation reaction rate is very limited ? The objective of this section is to address these questions.

On the other side, it must be kept in mind that  oxide growth in this range [1.5-4 nm] is strongly dependent on processing parameters like e.g., the pre-oxidation ramp and surface cleaning methods \cite{Baumvol99}. Moreover, the experimental  regime is relatively challenging for oxidation modeling. These two points emphasize the fact that only the  oxidation kinetics variation during the main oxidation stage can be discussed.  

\subsection{Comparison with Deal and Grove}

Our experimental kinetics data are compared with the two standard models.  Since our objective is to tackle the model limitations,  a continuous implementation of these models have been performed. The oxide thickness is obtained by integrating the following expression of the growth rate using very short timestep:

\begin{equation}
\frac{d X}{dt}= \frac{B}{2X+A}
\end{equation}

In the present case, for the diffusivity and the reaction rate, calibrated parameters of standard TCAD tool has been used \cite{Dios}. This set of parameters is supposed to give more accurate  results in the ultra-thin regime. Since, the oxidation temperature is below 1000$^{\circ}$C, the  linear reaction rate is govern by: 

\begin{equation}
\frac{B}{A}=1.25 \times 10^{05} \exp \bigg [ {-\frac{1.76 \ eV}{k_{B}T}} \bigg ] \quad  (nm/s)
\end{equation}

and the diffusivity 

\begin{equation}
B=2.833 \times  10^{09}  \exp \bigg [{-\frac{2.22 \ eV}{k_{B}T}} \bigg ] \quad (nm^{2}/s)
\end{equation}

\begin{figure}
\includegraphics[width=7cm]{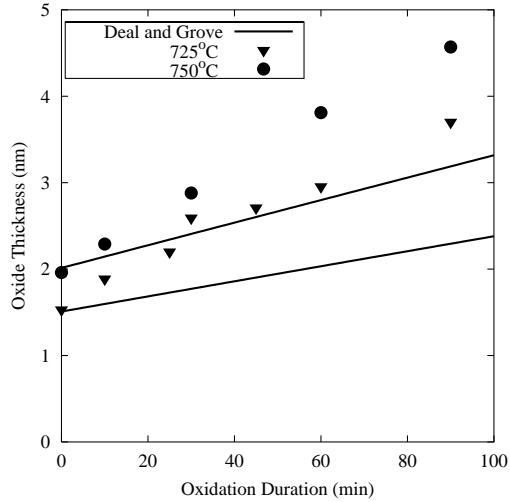}
\caption{Experimental kinetics at low temperature for 725$^{\circ}$C and 750$^{\circ}$C  (points) and the Deal and Grove model (lines). Oxide growth rate is underestimated leading to almost no oxide growth at 725$^{\circ}$C.
\label{fig:aps_dg}}
\end{figure}

Oxide growth has been simulated for the main oxidation steps at 725$^{\circ}$C and 750$^{\circ}$C. Since the initial oxidation ramp-up is not simulated, the theoretical curve must be  shifted upward to  match  the experimental thickness obtained at the end of the temperature ramp up. As shown in  figure \ref{fig:aps_dg},  the theoretical kinetic predicted by Deal and Grove \cite{deal65} significantly deports from experimental data. The oxide growth rate is strongly underestimated. Almost no oxide is created at low temperature, in contradiction with our oxidation experiments.

\subsection{Comparison with the Massoud model}

As previously discussed, the Massoud model is an extension of the Deal and Grove model to improve the description in the ultra-thin regime. In practice, the growth rate expression is corrected by adding a  supplementary term that exponentially decays with the oxide thickness: $C_{2}\exp (-\frac{X}{L})$ : 

\begin{equation}
\frac{d X}{dt}=\frac{B}{\displaystyle 2X+A} \times \big [\displaystyle 1+ C_{2}\exp(-\frac{X}{L})  \big]
\end{equation}

$C_{2}$ involves with an Arrhenius law:
\begin{equation}
C_{2}=61.52 \exp \bigg [{-\frac{2.56 \ eV}{k_{B}T}} \bigg ]  
\end{equation}

and the oxide thickness of this additional term is controlled by the parameter length (L=7 nm).\\

\begin{figure}
\includegraphics[width=7cm]{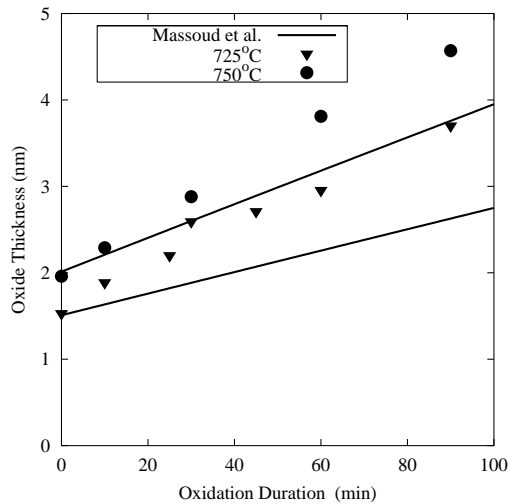}
\caption{Experimental kinetics at low temperature for 725$^{\circ}$C and 750$^{\circ}$C (points)  and the Massoud's model (lines). Oxide growth rate is clearly underestimated  at 725$^{\circ}$C.
\label{fig:aps_mass}}
\end{figure}

As shown in figure \ref{fig:aps_mass}, the introduction of the corrective term does not really improve the situation. Growth rate are still underestimated in  the lowest temperature regime. It is worth noting that the situation improves at  750$^{\circ}$C since it is much closer to  the lowest temperature (800$^{\circ}$C) used by Massoud et al. to  perform their calibration.\\

\subsection{Comparison with the reaction rate approach}

\begin{figure}
\includegraphics[width=7cm]{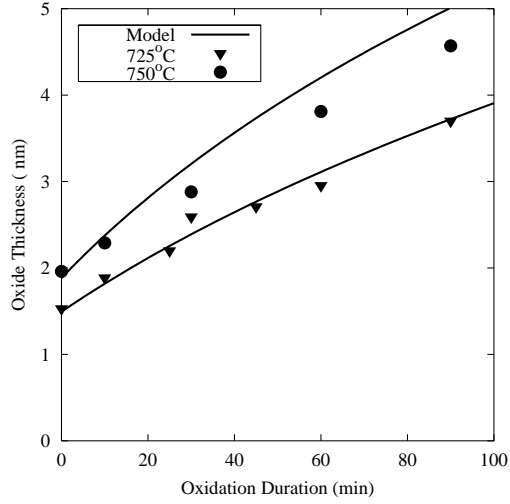}
\caption{Experimental kinetics at low temperature  for 725$^{\circ}$C and 750$^{\circ}$C (points) and our model derived by the reaction rate approach (lines). The first experimental kinetics is nicely described by the model whereas the second experimental kinetics is slightly overestimated.
\label{fig:aps_react}}
\end{figure}

A final comparison has been performed between the experimental results and the reaction rate approach. Results are shown in figure \ref{fig:aps_react}. The first stage of the oxidation process is not considered  since we are interested in simulating the main isothermal oxidation ramp.   The theoretical oxidation  kinetics are shifted upward in order to match the oxide thickness at the beginning of the oxidation ramp. Doing so, the agreement between the experimental kinetics and the model is  fine specially for the oxidation at 725$^{\circ}$C. 
 the range of oxide thickness predicted by the reaction rate approach is clearly relevant. For 750$^{\circ}$C, the model slightly overestimates  the growth rate but  the error   remains reasonable  (the maximum error is less  than 0.4 nm for the complete set of experimental data). Probably the most  interesting point is that the variation of the growth rate is relatively well defined. Moreover  the oxidation kinetics seems to be  not strictly linear as often  described by the two standard models (the reaction rate approach predicts a behavior in  $\propto$ t$^{0.8}$).

\section{Conclusion}

In conclusion, we  have proposed a combined experimental and theoretical study for the dry oxidation of silicon at low temperature. A model based on the reaction rate approach and on recent developments \cite{Almeida00} has been proposed. A set of parameters has been calibrated. A comparison with the two standard models shows a  very good predictivity both in the thin and the thick regime.  A complementary study  has been carried out to test the predictivity of the two standard models and of the reaction rate approach in the nanometric regime ([1.5-4 nm] range). Considering that  only two parameters have been calibrated, the results  obtained by this approach are very promising.  Further work is needed to refine and improve the model. For example, it would be of interest to test if  the reaction rate approach is able to describe the complex orientational effects observed experimentally by Irene et al. \cite{Irene87} or Ngau et al. \cite{Ngau02}. A further extension of this approach could also be  the  modeling of various oxynitridation processes  (NO, N$_{2}$O) used by the microelectronics industry.

\begin{acknowledgments}

The financial support of the European Union through the IST-2000-30129 FRENDTECH project (for the simulation work) and the IST-2000-26475 SODAMOS project (for ultra-thin gate oxidation experiments) is acknowledged. 

The authors would like to thank Jerzy Katcky from ITE (Varsaw) for the TEM analysis and Vincent Senez (IEMN) who initiated the work on oxidation modeling.  One of the authors (C. K.) is also strongly indebted to Peter Pichler (Franhofer FhG-IISB, Erlangen) for  support during the FRENDTECH project.
\end{acknowledgments}

\newpage



\begin{table}
\begin{tabular}{|c|c|c|c|}
\hline
T($^{\circ}$C)&n$^{0}_{O_{2}}$&$D$ (nm$^{2}$/s)&$K$ ($s^{-1}$)\\
\hline
800&1.50e-04 &27.5&7.0\\
\hline
850&1.43e-04&124.4&8.1\\
\hline
900&1.37e-04&355.6&15.8\\
\hline
950&1.38e-04&1035.9&23.8\\
\hline
1000&1.26e-04&2268.1&57.2\\
\hline
\end{tabular}
\caption{The value of the diffusivity ($D$) and of the reaction rate ($K$) obtained by the optimization procedure to match the experimental kinetics of Massoud et al. \cite{Massoud85a} at atmospheric pressure (P=1 atm).}
\label{table:oxid_one}
\end{table}

\begin{table}
\begin{tabular}{|c|c|c|}
\hline
Duration&Ellipsometry&TEM \\
\hline
10 Min.& 18.90  $\pm$ 0.3  \AA  & 19  $\pm$ 1.0 \AA \\
30 Min.& 27.10  $\pm$ 0.3  \AA  & 27  $\pm$ 1.0 \AA \\
\hline
\end{tabular}
\caption{Validation of our ellipsometry measurements by TEM analysis on two samples oxidized at 725$^{\circ}$C  with the same pre-oxidation ambient N$_{2}$/O$_{2}$.}
\label{table:oxid_two}
\end{table}

\cleardoublepage

\end{document}